\documentclass[]{piparticle-final}
\usepackage{graphicx}
\usepackage{amsmath}

\usepackage{cite} 
\usepackage{epstopdf}

\begin{document}

\volume{6}               
\articlenumber{060014}   
\journalyear{2014}       
\editor{L. A. Pugnaloni}   
\reviewers{K. To, Institute of Physics, Academia Sinica, Taipei, Taiwan.}  
\received{21 November 2014}     
\accepted{5 December 2014}   
\runningauthor{I. Zuriguel}  
\doi{060014}         

\title{Invited review: Clogging of granular materials in bottlenecks}

\author{Iker Zuriguel\cite{inst1}\thanks{E-mail: iker@unav.es}}

\pipabstract{
During the past decades, notable improvements have been achieved in the understanding of static and dynamic properties of granular materials, giving rise to appealing new concepts like jamming, force chains, non-local rheology or the inertial number. The `saltcellar' can be seen as a canonical example of the characteristic features displayed by granular materials: an apparently smooth flow is interrupted by the formation of a mesoscopic structure (arch) above the outlet that causes a quick dissipation of all the kinetic energy within the system. In this manuscript, I will give an overview of this field paying special attention to the features of statistical distributions appearing in the clogging and unclogging processes. These distributions are essential to understand the problem and allow subsequent study of topics such as the influence of particle shape, the structure of the clogging arches and the possible existence of a critical outlet size above which the outpouring will never stop. I shall finally 
offer 
some hints about general ideas that can be explored in the next few years.}

\maketitle

\blfootnote{
\begin{theaffiliation}{99}
   \institution{inst1} Departamento de F\'{\i}sica, Facultad de Ciencias, Universidad de Navarra, 31080 Pamplona, Spain.
\end{theaffiliation}
}

\section{Clogging in bottlenecks, a multiscale and multidisciplinary problem.}

When a system of discrete bodies passes through a constriction, the interactions among the particles might lead to the development of clogging structures that, eventually, completely arrest the flow. This phenomenon is observed in a wide range of systems with relevant consequences. Clogging of granular materials in silos may force a production line to be stopped. Much in the same way, clogging of suspended hydrated particles is a major issue concerning oil and gas transport through pipelines \cite{Sloan}. At a smaller spatial scale, clogging leads to intermittent flow when a dense suspension of colloidal particles passes through a constriction in a microchannel \cite{Haw,Genovese}. A straightforward application of the understanding that could be gained concerning clogging in suspensions is found in ecological engineering. Nowadays, an alternative that is becoming widely used for removing pollutants from wastewater is the use of subsurface flow treatment. The most 
important drawback of this technique is its 
unpredictable lifetime, mostly limited by clogs that obstruct the pores \cite{Knowles}. At an even smaller scale, intermittent flows are observed when electrons on the liquid helium surface pass through nanoconstrictions \cite{Rees}. Finally, clogging can also develop when crowds in panic are evacuated through emergency exits that cannot absorb the amount of people approaching the doors \cite{HelbingNature,Helbing_transpscience,Moussaid}.

All these examples of clogging take place in a broad range of systems where widely different forces are at play: those concerning the interactions among particles as well as those related to the interaction between particles and the surrounding media. For the case of non-cohesive inert grains, gravity and contact forces are the only relevant ones. For particle suspensions, however, the hydrodynamics of the flowing fluid as well as the capillary effects must be taken into account. Dynamics of crowds through bottlenecks are even more difficult to approach theoretically, yet a social force model has been proved to adequately reproduce the observed behavior in some circumstances.

In the last decade, the number of works published about clogging has experienced a sudden increase, which constitutes a gauge of the interest and relevance of this phenomenon. Despite this, the physical mechanisms behind clogging are still not well understood. Several issues contribute to this, but probably the most important one concerns the local character of clogging when compared, for example, with the global nature of jamming \cite{Liu}, a scenario that has attracted much more attention over the last years. This `local character of clogging' seems to complicate the definition of global extensive variables within the system which could be used to characterize the phenomenology.

In this manuscript I shall give a brief summary of the major advances achieved in the understanding of clogging of non-cohesive inert grains at a bottleneck. I will start by presenting results on the static silo, then I will move to the case of a vibrated silo, and finally I will present some conclusions and mention several questions that ---from my point of view--- should be addressed in the forthcoming years.

\section{Clogging in hoppers or silos}
\label{clogging}

The development of clogs that obstruct the flow in the discharge of bins or silos by gravity is a problem that has always worried the engineering community \cite{Kvapil,Walker,Sakaguchi,Drescher}. The goal in all these works was finding a ratio of the outlet to particle size that could guarantee the absence of clogging. For non-cohesive materials, it was known that this value is around $5$ although, depending on the particle properties, it could increase up to $10$. Nevertheless, little was understood about the mechanism that triggers a clog and the physical variables that control its development. Indeed, it was not until the beginning of this century when the scientific community started to carefully investigate this problem \cite{To2001}.

\subsection{Avalanche size distribution}

\begin{figure}
\begin{center}
\includegraphics[width=0.5\textwidth]{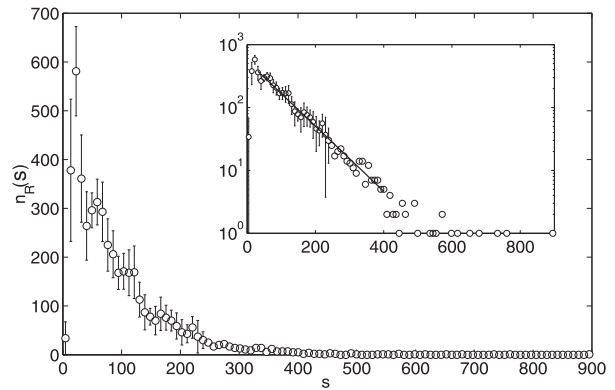}
\end{center}
\caption{Histogram $n_R(s)$ for the number of grains $s$ that flow between two successive clogs. Data correspond to a circular orifice of $6\;mm$ diameter and glass beads with
a diameter of $2\;mm$. More than 4000 events were recorded. In the inset, a semilogarithmic plot with the solid line indicating an exponential fit.} \label{figure1}
\end{figure}

One of the first questions that was tackled concerns the statistics of the avalanche sizes (the usual measured magnitude is the number of grains that flow out of the silo from the breakage of a clogging arch, until the development of a new one). The key feature of avalanches, which is now well accepted, is that the distribution of their sizes follows an exponential decay (see Fig. \ref{figure1}), a result reported for the first time in \cite{Clement}. This trend was explained in \cite{Zuriguel2003} by assuming that the probability of clogging is constant during the whole avalanche, a behavior observed for all the outlet sizes explored. Afterward, a many-particle-inspired theory was proposed based on a continuity equation in polar coordinates \cite{Helbingexponential}. Although friction, force networks or inelastic collapse of the particles were not taken into account, the authors reproduced the exponential decay of the avalanche (or burst) sizes. The 
intermittent flows reported were 
explained in terms of a random alternation between particle propagation and gap propagation. Very recently, a probabilistic model ---in which the arches were modeled by a one-dimensional stochastic cellular automaton--- also sheds light on the origin of the exponential character of the avalanche size distribution \cite{Masuda}. The stochastic nature of the clogging process was also suggested in \cite{bob} where it was shown that the particles that end up forming the clog were totally uncorrelated at the beginning of the avalanche.

The exponential decay of the avalanche size distribution has been reported in several arrangements: 2D and 3D silos \cite{Zuriguel2003,Janda2008,Perez}, 2D hoppers \cite{To2005,kondic}, 2D and 3D tilted hoppers and silos \cite{Sheldon,Thomas}, silos with the presence of obstacles \cite{Zuriguelobstacle,Zuriguelobstacle2}, 2D silos where the particles were driven by different gravity forces \cite{Arevalo}, and fluid driven particles in 2D and 3D \cite{Guariguata,Lafond}. However, there are some examples where this exponential tail breaks down. Those situations are typically related to a breaking of the symmetry of the problem as in the following cases: 1) usage of particles with shapes that are not spherical \cite{Zuriguel2005}; 2) emplacement of multiple orifices \cite{Mondal}; 3) implementation of slots in 3D silos instead of the normal circular orifices \cite{Franklin}. In the latter case, a power law distribution was observed as it will be explained in section \ref{Orifice geometry} 
Incidentally, 
power law distributions were also numerically obtained when considering internal avalanches, defined as the number of grains that move inside the silo between consecutive clogs. This result was compatible with the idea of Self Organized Criticality, in analogy with avalanches developed at the surface of a pile \cite{Manna}. This work was, indeed, one of the first approaches to the avalanche statistics in the silo problem.

\subsection{Does a critical outlet size exist?}

\begin{figure}
\begin{center}
\includegraphics[width=0.52\textwidth]{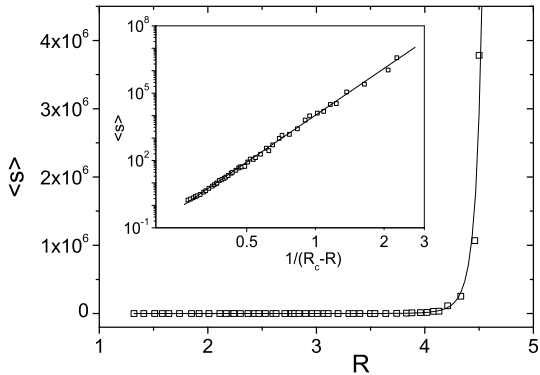}
\end{center}
\caption{Mean avalanche size $\langle s \rangle$ vs. $R$, the ratio between the outlet and particle diameters. The solid line is
a fit with the equation: $\langle s \rangle = A (R_C-R)^{-\gamma}$; with $R_C=4.94\pm0.03$, $\gamma = 6.9 \pm 0.2$, and $A = 9900 \pm 100$. Inset: mean avalanche size $\langle s \rangle$ vs.  $1/(R_C-R)$. Note the logarithmic scale. Figure reprinted with permission from Ref. \cite{Zuriguel2005}. Copyright (2005) by the American Physical Society \cite{copyright}.} \label{figure2}
\end{figure}

\begin{figure}
\begin{center}
\includegraphics[width=0.45\textwidth]{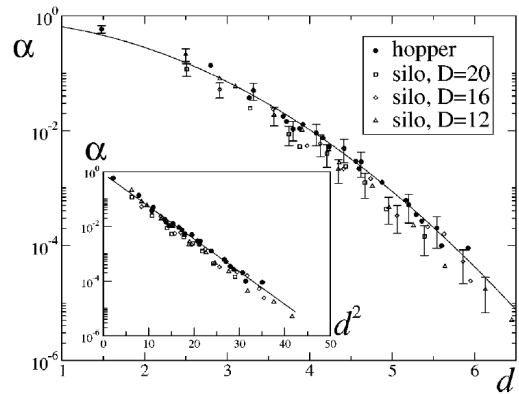}
\end{center}
\caption{Variation of the decay rate $\alpha$ with hopper or silo exit $d$. $\alpha$ is obtained from the fittings of the avalanche sizes with $F(s)=e^{-\alpha (s-s_0)}$. The solid line is the fitted curve $\alpha=A e^{-Bd^2}$ with $A=0.846$ and $B=0.275$. The inset shows the same data and the fitted curve with $d^2$ plotted in the x axis. Figure reprinted with permission from Ref. \cite{To2005}. Copyright (2005) by the American Physical Society \cite{copyright}.} \label{figure3}
\end{figure}

Considering the exponential character of the avalanche size distribution, its first moment (the average avalanche size $\langle s \rangle$) can be easily calculated and used to study the dependence of clogging on the size ratio between the outlet and the particles. For spherical beads in a 3D silo, a divergence of the avalanche size was reported for an outlet diameter about $5$ times the bead diameter (see Fig. \ref{figure2}) \cite{Zuriguel2005}. This divergence was shown to be robust, as it holds for particles with widely different properties. Among these, the shape of the particle was reported to be the most influential on the critical outlet size value. Nevertheless, in a subsequent work, the existence of such a critical outlet size was challenged by K. To \cite{To2005}. In a two dimensional silo, it was shown that several empirical fits agree reasonably well with the experimental data: some were compatible with the existence of a critical outlet, but others were not (see 
Fig. \ref{figure3}).

Following this idea, Janda et al. \cite{Janda2008} demonstrated that one of the non-divergent expressions proposed in \cite{To2005} could be analytically deduced using both, the probability that a given number of particles meet above the outlet ---as suggested by Roussel et al. \cite{Roussel}--- and the probability of finding arches of a given size within a granular deposit ---as found in \cite{Arevaloarches,Pugnaloniarches}. Unfortunately, the reasoning used in two dimensions was not applicable to three dimensional silos, where the transition seems to actually exist. Interestingly, this clogging transition has been also identified for inclined silos and orifices \cite{Sheldon,Thomas}, as well as in the discharge of granular piles through an orifice below its apex \cite{Atman}. Very recently, the mean avalanche size has been put on relation with the fraction of clogging configurations that are sampled by the orifice, suggesting that $\langle s \rangle$ should increase exponentially with the hole width raised 
to 
the system dimensionality \cite{Thomasarxiv}. According to these results, clogging is akin to the jamming and glass transitions in the sense that there is not any sharp discontinuity in the behavior, but a dramatic increase of the relaxation times as the orifice size is enlarged.

\subsection{Clogging arches}

\begin{figure}
\begin{center}
\includegraphics[width=0.4\textwidth]{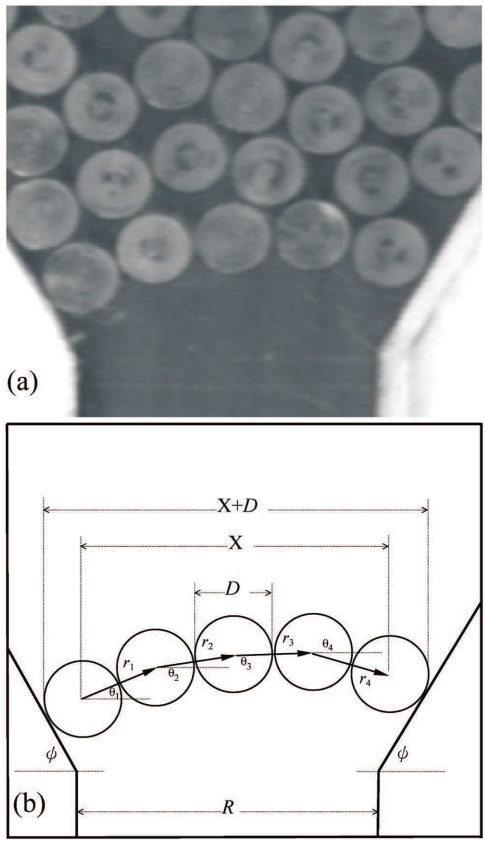}
\end{center}
\caption{(a) Image of a typical clogging arch. (b) Configuration of the arch where $r_i$ illustrate the $i$ steps of the restricted random walk model proposed in \cite{To2001}. Figure reprinted with permission from Ref. \cite{To2001}. Copyright (2001) by the American Physical Society \cite{copyright}.} \label{figure4}
\end{figure}

Complementary to the analysis of the avalanche sizes, some authors have paid special attention to the arches that clog the orifices. Clogging arches are structures of several mutually stabilizing particles that have to span, at least, the size of the constriction. In his seminal work, To et al. introduced a simple model to explain the clogging probability based in the geometry of the clogging arches (see Fig. \ref{figure4}) \cite{To2001}. They proposed that the position of particles in a clogging arch is the result of a random walk model with some restrictions: 1) the horizontal span of the arch should be larger than the orifice; 2) the arch has to be convex everywhere; 3) particles conforming the arch should be in contact with each other. This model nicely reproduced the clogging probability for hopper angles below $75^{\circ}$. In a subsequent work \cite{To2002}, the same authors introduced an approximation of the arch 
shape to a circular arc centered at the apex of the hopper cone. From this, they calculated 
detailed properties of the clogging arches, such as the number of disks conforming them, finding good agreement with the experimental results. Some of the ideas proposed by To et al. were corroborated in Ref. \cite{Garcimartin} where it was reported that the aspect ratio of the arches (the height divided by half the span) tends to one, a result that is compatible with a semicircular shape. In addition, it was shown that the convexity condition assumed by To is not necessarily fulfilled in all the particles. Indeed, $17\%$ of the particles had an associated angle with their two neighbors above $180^{\circ}$. Hence, arches were locally concave at that particle, a situation which was named `defect'. Despite this seemingly mismatch with the restricted random walk model, a strong inverse correlation of the angle associated to a particle and the one of their neighbors was also shown. Apparently, this inverse correlation 
compensates the apparition of defects and preserves the validity of the 
restricted random walk model.

Apart from the works mentioned above, where only the geometry of the arches was evaluated, there have been preliminary attempts to consider the forces involved within the particles conforming the arches. In \cite{Longjas}, force analysis was used to calculate the jamming probability of mixed sizes disks that move downwards under gravity in a two-dimensional hopper. The authors focused in the simplest case of arches formed by three discs which, for the outlet size employed, were the most common. Finally, Hidalgo et al. \cite{hidalgo} performed numerical simulations and found that ---in clogging arches--- the tangential forces in the `defects' were very high, while normal forces were abnormally low. The outcome concerning tangential forces is somehow expected as friction is necessary to stabilize defects. On the contrary, the result concerning the normal forces is rather counterintuitive, but it was in accord with a previous forecasted prediction based on 
experimental works on the stability of arches in a 
vibrated silo \cite{Lozanovibrated}.

\subsection{Orifice geometry.}
\label{Orifice geometry}

As stated above, clogging is a local phenomenon in the sense that it always takes place at the constriction. According to this, it seems rather obvious that the properties of the confining geometry would importantly affect the clogging process. An evidence of this can be found in \cite{To2001} where it is shown that the clogging probability in a hopper is notably reduced as the hopper angle increases from $60^{\circ}$ to $75^{\circ}$. On the contrary, hopper angles below $60^{\circ}$ give rise to similar clogging probabilities. The reason for this seems to be founded on the fact that, for sufficiently flat hoppers, the grains develop a  spontaneous internal angle of repose which acts as an internal hopper.

The effect of this internal angle of repose is also relevant in the works of Durian's group who implemented inclined orifices and silos \cite{Sheldon,Thomas}. This practice is relatively common in industrial hoppers and hence, knowing the way in which clogging is affected, becomes significant. The authors have proved that increasing the tilting angle of the orifice or silo augments the propensity to clog according to a reduction in the projection of the aperture area against the average flow direction. In addition, a clogging phase diagram is proposed combining tilting angle and outlet size. For circular apertures, the same diagram is found for four grain types (including prolate and oblate ones). For slots, however, the shape of the phase diagram for the case of lentils and rice seems to be different than for more isotropic grains, an effect attributed to an alignment between grains and slit axes.

The use of slots instead of circular orifices was already proved to be beneficial to prevent clogging \cite{Davies}. In this work, some conservative guidelines are given to select the minimum outlet size that assures no flow interruption. While for horizontal and vertical slots the ratio of slot width to particles size are $3.3$ and $4.6$ respectively, for horizontal circular outlets the ratio of orifice diameter to particle size is $6.4$. This number may seem considerably larger than the ones reported in \cite{Zuriguel2005}, but it should be taken into account that particles of different properties (including anisotropic ones) were employed. Even more importantly, in a subsequent work, it was reported that, as the length of the slot increases, the avalanche size distribution departs from the exponential behavior displaying a power law decay \cite{Franklin}. This behavior is explained in terms of a model where a slot is represented by a series of statistically independent 
cells whose length is related with a hypothetical distance 
along which particles' movement is correlated. Interestingly, the model matches experimental outcomes for a correlation distance of around $10$ particle diameters. Nevertheless, this result needs to be confirmed as in other experiments using slots, the avalanche size distribution has been found to be exponential for different types of grains \cite{Thomas}.

A configuration which is closely related to the slot geometry is the placement of several aligned orifices. Very recently it has been reported that clogging can be significantly reduced by having more than one exit orifice. In this situation, when one of the orifices jams, the flow through the adjacent unjammed orifice might cause perturbations in the clogging arch, destroying it and leading to a sequence of jamming and unjamming events \cite{Kunte}. The necessary condition to observe this behavior is, of course, that orifices are close enough to each other. In the same line, Mondal and Sharma \cite{Mondal} have shown that adjacent outlets start affecting each other when the distance is approximately three times the diameter of the particles. These authors point toward the importance of stable particles (adjacent to the arches) resting on the base of the silo. Remarkably, the role of these particles was in fact overseen in previous works that analyzed the properties of clogging arches \cite{To2001,To2002,
Garcimartin,Longjas,hidalgo,Lozanovibrated}.

A smart alternative to alter the clogging process consists of placing an obstacle just above the outlet. In \cite{Zuriguelobstacle}, it was reported that the clogging probability may be reduced up to $100$ times if the obstacle position is properly selected. This dramatic effect was attributed to a reduction in the pressure (or particle confinement) in the orifice neighborhood, which apparently favors arch destabilization. The explanation given is that particles colliding above the orifice ---which eventually could form a clogging arch--- are not easily stabilized if there is not a certain confinement that facilitates energy dissipation. This idea was supported by the observation of a sudden increase on the number of particles ejected upwards in the outlet proximities when the obstacle was placed. In the same work, simulations of a silo filled with a few layers of grains revealed the same kind of clogging reduction as the layer of grains above the orifice was reduced, then 
confirming the important role of 
pressure in the process. In a subsequent work \cite{Zuriguelobstacle2}, it was shown that the effect of the obstacle is enhanced as the outlet size enlarges. It is noteworthy that, in all the cases, the clogging reduction is achieved with just a tiny alteration of the flow rate (up to $10\%$ in the worst situation). In these works, an issue that remains unclear is the role of the packing fraction above the orifice. Clearly, the placement of the obstacle affects this variable which should be, indeed, related to pressure. Nevertheless, robust measurements of volume fraction are extremely difficult near the outlet due to the existence of strong gradients.

As far as I know, there has been only one attempt to unveil the role of volume fraction on clogging in dry granular media \cite{Unac}. In this work, a pseudo-dynamic model was implemented to prepare samples with different initial configurations by means of a tapping procedure. Although packing fraction affects clogging, their main conclusion is that this is not a good macroscopic parameter to predict the size of the avalanches that would flow through a given aperture, suggesting that further information about the packing properties is necessary. A nice alternative to study the effect of packing fraction on the ability of a system to develop clogs is the use of solid particles suspended in a fluid. This is precisely what it was done in \cite{Valdes2} where it was proved that the probability of bridge formation increased with the volume fraction. Note that this system has the advantage of allowing a better control of the volume fraction than just varying the initial configuration as done in \cite{Unac}.

\subsection{Effect of polydispersity and particle shape}

A recursive topic that arises in the granular community is the roles that size polydispersity and particle shape play on the behavior of such materials. For the case of clogging in silos, in \cite{Zuriguel2005} it was reported that polydisperse samples displayed the same exponential decay of the avalanche size than monodisperse ones. Furthermore, it was revealed that polydispersity had a negligible effect in the critical outlet size above which clogging would not occur. In \cite{Pournin}, clogging of bidisperse samples was also shown to be similar to the monodisperse case as long as segregation is prevented. In addition, the authors propose that the parameter that should be considered to characterize the mixture is the particles volume-average diameter.

Contrary to polydispersity, particle shape seems to play a major role in clogging development as evidenced using prolate (rice) and oblate (lentils) particles \cite{Zuriguel2005}. The critical outlet size increases (i.e., clogging is more likely) when anisotropic particles are employed, a result coherent with that obtained in fluid driven suspensions of mica flakes when compared with glass beads \cite{Valdes2}. An issue that is still open concerning anisotropic particles in the discharge of a silo is the characteristic particle length that should be chosen to compare with that of the orifice. In addition, there is a lack of experiments or simulations about the effect of using faceted particles in the clogging probability. Some words have been written suggesting that faceted particles dramatically increase the clogging ability due to their tendency to align \cite{EPJETanzaki,Hohner}, yet there are not systematic results on this interesting topic.

\subsection{Dynamic signatures of clogging}
\label{dynamic}

\begin{figure}
\begin{center}
\includegraphics[width=0.45\textwidth]{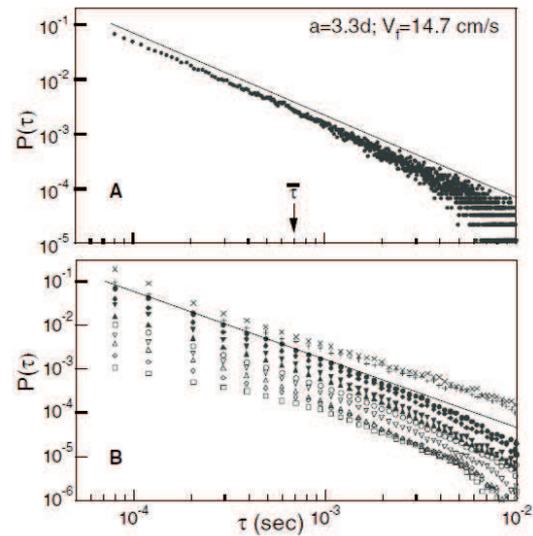}
\end{center}
\caption{(A) Probability distributions, $P(\tau)$, of the time intervals
$\tau$ between collisions, on a log-log scale for an outlet length $3.3$ times the diameter of the particles. The solid line
corresponds to a power law $P(\tau) \sim \tau ^{-3/2}$. The average time
interval between impulses is marked in the figure. (B) $P(\tau)$
on a log-log scale for different opening sizes ranging from $3$ to $16$ times the diameter of the particles (the curves on the top correspond to the smaller outlet sizes). The curves are displaced vertically for clarity. The solid line is the power law:
$P(\tau) \sim \tau ^{-3/2}$. Figure reprinted with permission from Ref. \cite{Menon}. Copyright (2002) by the American Physical Society \cite{copyright}.} \label{figure5}
\end{figure}

Provided that clogging in bottlenecks is a consequence of the sudden formation of a stable arch at the very narrowing, it is a big challenge to find dynamical descriptors in the flowing state that can be used to predict an eventual arrest of the flow. The first important result about this challenge was reported by Longhi et al. \cite{Menon} who studied the impulses recorded by a force transducer at the hopper boundary near the orifice. Although the distribution of impulses does not reveal any static signature of jamming, the distribution of the time intervals between collisions ($\tau$) produces interesting distinctive features as the outlet size is reduced and approaches the clogging region (see Fig. \ref{figure5}). In fact, this distribution tends to a power-law $P(\tau)\sim \tau^{-3/2}$ implying that the mean time interval tends to diverge as the outlet is reduced. This is so even when the average time computed from a finite (albeit large) data set shows a relatively negligible dependence on the outlet 
size.

In \cite{Janda_fluctus}, the flow rate properties were carefully examined in a two-dimensional silo for outlet sizes both, above and below the supposed critical outlet size. Even though the average flow rate behaved smoothly and did not display any characteristic property near the critical size, it was observed that the flow rate fluctuations are non symmetric for small apertures. For large orifices, the measurements of the instantaneous flow rate ($q$) display Gaussian-like distribution of the fluctuations around the average. Nevertheless, as the outlet size is reduced, temporal interruptions of the flow are evidenced by the development of a peak at $q=0$ in addition to the one that corresponds to the flowing regime.

In this direction, a step further was performed by Tewari et al. \cite{bulbul} who implemented event-driven simulations to analyze the velocity fluctuations of grains flowing through a hopper. The analysis in this work was not restricted to the region of the orifice, as all the grains of the silo were studied. Interestingly, although the kinetic temperatures are always higher at the boundaries of the silo, the correlation times display a tendency that reverses as the outlet size is reduced: whereas for high flow rates (far above the critical outlet size) the flow at the center has longer autocorrelation times than at the boundary, the opposite is valid for low flow rates as fluctuations relax more slowly at the boundaries. In this work, it is also suggested that clogging is preceded by the appearance of vortices that nucleate at the corners of the hopper and extend inwards.

\section{Vibrated silos: clogging and unclogging}
\label{clogandunclog}

\begin{figure}
\begin{center}
\includegraphics[width=0.45\textwidth]{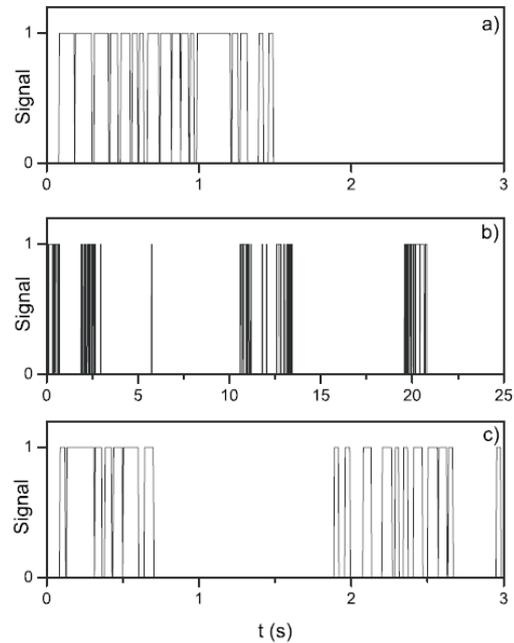}
\end{center}
\caption{Signal from a photosensor at the exit of a silo: a value of $1$
indicates that a particle is blocking the beam, zero means that the
beam is unobstructed. (a) Static silo. (b) Vibrated silo. (c) A
zoom of the signal shown in (b) during the first three seconds, the
same time stretch as in (a). All the data were obtained
using an orifice of diameter $3.05$ times the beads diameter. Figure reprinted with permission from Ref. \cite{Mankocvibrated}. Copyright (2009) by the American Physical Society \cite{copyright}.} \label{figure6}
\end{figure}

Up to now, I have described investigations related to the clogging process presuming that, once a clogging bridge is formed, all the kinetic energy is dissipated and the structure is forever stable. Nevertheless, an alternative approach can be implemented, which consists on applying an external input of energy and study its effect on clogging. This strategy gives rise to a dramatic change in the observed dynamics when the orifice is small, i.e., in the region where clogging is frequent. Unlike the case of a static silo, the flow in the vibrated silo is characterized by the alternation of jamming and unjamming events. Indeed, apart from the flow rate fluctuations found in a static silo \cite{Janda_fluctus}, in the vibrated case long flow interruptions were present. These were attributed to arches that form and were initially stable, but destabilize as a consequence of vibrations (see Fig. \ref{figure6}). This behavior suggests that the intermittent flow in vibrated silos 
can be split in two 
different, independent processes: clogging and unclogging. Following this line of reasoning, Mankoc et al. \cite{Mankocvibrated} reported that the probability that a system clogs does not depend on the vibration, which only introduces a non-zero probability of unclogging once an arch has blocked the orifice. This probability of unclogging was measured in three different ways which led to consistent results whose most conspicuous feature was an increase of the unclogging probability with the outlet size.

\begin{figure}
\begin{center}
\includegraphics[width=0.45\textwidth]{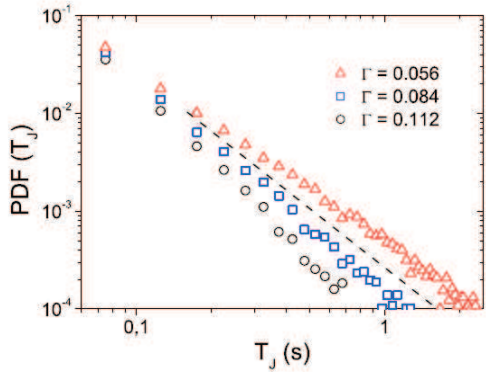}
\end{center}
\caption{Histogram, in logarithmic scale, for the time lapses that the
orifice remains blocked in a vibrated silo for different vibration accelerations as indicated in the legend. Data correspond to an outlet size
$1.78$ times the particle diameter. The dashed line has a slope of two evidencing that, for the smallest acceleration displayed, the slope is smaller than two. Figure reprinted with permission from \cite{Jandavibrated}. Copyright (2009) by IOP.} \label{figure7}
\end{figure}

Janda et al. \cite{Jandavibrated} devised a similar experiment in which the hopper wall of an eccentrically discharged silo was a piezoelectric, allowing a local perturbation of the clogging arch. In this sense, this work is conceptually different than that of Mankoc et al. where the whole silo was vibrated. The most interesting result revealed by Janda et al. was that the distribution of times that the system takes to get unclogged exhibits a power law decay. At low vibration accelerations,  anomalous statistics for the jamming times were evidenced as the exponent $\alpha$ of the power law was below $2$ and the first moment could not be calculated (see Fig. \ref{figure7}). This property is, indeed, strongly reminiscent of the anomalous dynamics usually observed for creeping flows of glassy materials. In a recent work, this behavior has been shown to be universal in other systems of macroscopic particles flowing through a bottleneck like sheep, a model of pedestrians, and 
colloids \cite{Zuriguel_
phasetransition}. Furthermore, for the case of inert grains, several variables have been shown to affect the value of the exponent going from $\alpha>2$ to $\alpha \leq 2$, i.e., from an unclogged situation (where averages can be defined) to a clogged scenario (where the average flow rate would tend to zero as the measuring time increases). These variables are: the intensity of vibration, the outlet size, the height of the layer of grains above the outlet, and the inclination of the 2D silo with respect to the vertical which modifies the component of the gravity affecting the grains. Increasing the intensity of vibration and enlarging the outlet size favors the development of unclogged situations, while increasing the layer of grains or the silo verticality facilitates the transition to clogging. A similar idea was anticipated by Vald\'es and Santamarina \cite{Valdes1} who suggested that the acceleration that would be required for unclogging increases with increasing skeletal forces 
in the particles
forming the bridge. Furthermore, they related this prediction to the higher stability exhibited by the arches formed in a suspension when subjected to high fluid
velocities \cite{Muecke}.

Finally, in a 2D vibrated silo which allowed observation of the clogging structures, it was established a relationship among the bridge geometry and its resistance to vibration \cite{Lozanovibrated}. In particular, it was revealed that the intensity of vibration at which the arches collapse is inversely correlated with the maximum angle among the particles conforming it. For the particular case of angles above $180^0$ (the so called defects), this dependence was explained in terms of a very simple force analysis. In summary, from this work it was concluded that arches break at defects and, the larger the maximum angle, the weaker is the arch.

\section{Perspectives}

After more than a decade of research, significant advance has been achieved in the understanding of clogging. Despite all that, the relevance of the remaining open issues and the importance of the consequences that clogging has from an applied point of view, hint about an augment of activity on this topic in the forthcoming years. Probably, a sensible approach that should be investigated is isolating the dynamic and geometric contributions in the development of clogs. Effectively, a clogging arch should have a structure compatible with the confined geometry. But in addition, this structure has to be able to persist until all the kinetic energy of the system is dissipated. Unfortunately, increasing the outlet size leads to a modification of the geometry of the problem (as the span of clogging arches has to increase), but also affects to the velocity of the particles (which increases with the square root of the orifice diameter). In a recent work, Ar\'evalo et al. made initial progress in this direction by 
exploring clogging when reducing the driving force up to $10^{-3} g$ where $g$ is the gravity; but undoubtedly, new strategies should be devised to understand the effect of dynamics in the clogging process.

A situation which seems simpler as dynamic effects are removed is the study of unclogging as explained in section \ref{clogandunclog} The power law decays observed in the time that the system needs to become unclogged, suggest a creeping process where the bridges would age with time, increasing their endurance. A straightforward way of testing the validity of these ideas would be an analysis of this process using photoelastic particles to evaluate temporal evolution of the forces within the arch \cite{bob2}. The usage of this kind of particles could also be implemented in order to unveil an old question concerning the relationship among clogging arches and force chains.

Another issue that remains unsolved is, whether or not, clogging can be seen as a phase transition and, if so, what kind of transition clogging is. As explained above, from the measurements of unclogging times in a vibrated silo, a divergence has been found that can be used to rigourously characterize the clogged state through the definition of a `flowing parameter' \cite{Zuriguel_phasetransition}. A thorough inspection of the dependence of this parameter on different variables becomes necessary to corroborate its usefulness. Nonetheless, as this approach is based on the unclogging times, it cannot be used for the `singular' case of a static silo where, if formed, clogs last forever. In such scenario, instead of the traditional way of studying the divergence of the avalanche size as the outlet is enlarged, I believe that it is pertinent to approach the transition from the flowing region. There, it should exist some parameter that reveals distinctive behavior when the outlet size is reduced as explained 
in 
section \ref{clogging}\ref{dynamic} A reminiscent problem of this alternative is the difficulty of choosing a region where to perform the analysis as the silo is precisely characterized by the existence of strong spatial and temporal gradients. In this sense, a geometry that becomes promising is a narrow pipe without any constriction where clogs may develop at any place \cite{Hadjigeorgiou,Tsai,Jandapipe}. Apart from a zone at the top of the pipe where the pressure increases with depth, a rather homogeneous behavior should be observed within the rest of the system, allowing clean measurements of variables like velocity, density and so on. The study of this geometry can also be seen as an intermediate stage between clogging and jamming as it is also the case of `jamming by pinning' \cite{Reichhardt} where the increase of the number of obstacles in the system (and so the characteristic distance between them) was shown to reduce the density at which the system jams.

%
%

\begin{acknowledgements}
I would like to thank the referee Kiwing To whose comments have, undoubtedly, helped to improve the quality of this manuscript. I am very grateful to Angel Garcimart\'in, Diego Maza, Carlos P\'erez-Garc\'ia and Luis Pugnaloni, without whom this work would never have been possible.
\end{acknowledgements}


\begin{thebibliography}{50}
\bibitem{copyright} Readers may view, browse, and/or download material for temporary copying purposes only, provided these uses are for noncommercial personal purposes. Except as provided by law, this material may not be further reproduced, distributed, transmitted, modified, adapted, performed, displayed, published, or sold in whole or part, without prior written permission from the American Physical Society.
\end{thebibliography}
\end{document}